\RequirePackage{ifpdf}
\ifpdf % We are running pdfTeX in pdf mode
\documentclass[pdftex]{sigma}
\else
\documentclass{sigma}
\fi

\def\D{\mathbb{D}}
\def\T{\mathbb{T}}

\def\G{\mathbb{G}}

\numberwithin{equation}{section}

\begin{document}

\allowdisplaybreaks

\renewcommand{\thefootnote}{$\star$}

\renewcommand{\PaperNumber}{028}

\FirstPageHeading

\ShortArticleName{Dynamical Studies of Equations from the Gambier Family}

\ArticleName{Dynamical Studies of Equations\\ from the Gambier Family\footnote{This
paper is a contribution to the Proceedings of the Conference ``Integrable Systems and Geomet\-ry'' (August 12--17, 2010, Pondicherry University, Puducherry, India). The full collection is available at \href{http://www.emis.de/journals/SIGMA/ISG2010.html}{http://www.emis.de/journals/SIGMA/ISG2010.html}}}

\Author{Partha GUHA~$^\dag$, Anindya GHOSE CHOUDHURY~$^\ddag$ and Basil GRAMMATICOS~$^\S$}
\AuthorNameForHeading{P.~Guha, A.~Ghose Choudhury  and B.~Grammaticos}

\Address{$^\dag$~S.N. Bose National Centre for Basic Sciences,\\
\hphantom{$^\dag$}~JD Block, Sector-3, Salt Lake, Calcutta-700098, India}
\EmailD{\href{mailto:partha@bose.res.in}{partha@bose.res.in}}

\Address{$^\ddag$~Department of Physics, Surendranath  College,\\
\hphantom{$^\ddag$}~24/2 Mahatma Gandhi Road, Calcutta-700009, India}
\EmailD{\href{mailto:a_ghosechoudhury@rediffmail.com}{a\_ghosechoudhury@rediffmail.com}}
\Address{$^\S$~IMNC, Universit\'e Paris VII-Paris XI, CNRS, UMR 8165, Bât. 104, 91406 Orsay, France}
\EmailD{\href{mailto:grammaticos@univ-paris-diderot.fr}{grammaticos@univ-paris-diderot.fr}}

\ArticleDates{Received December 10, 2010, in f\/inal form March 17, 2011;  Published online March 22, 2011}

\Abstract{We consider the hierarchy of higher-order Riccati equations and
establish their connection with the Gambier equation. Moreover we
investigate the relation of equations of the Gambier family to
other nonlinear dif\/ferential systems. In particular we explore
their connection to the generalized Ermakov--Pinney and
Milne--Pinney equations. In addition we investigate the consequence
of introducing Okamoto's folding transformation which maps the
reduced Gambier equation to a Li\'{e}nard type equation. Finally
the conjugate Hamiltonian aspects of certain equations belonging
to this family and their connection with superintegrability are
explored.}


\Classification{34C20; 70H05}

\section{Introduction}

The Gambier equation plays, for linearizable systems, the same
role as the Painlev\'e~VI. Just as one can derive, starting from
the latter, all the remaining Painlev\'e equations through
appropriate limiting procedures, in the same way the Gambier
equation is a ``master'' system for the linearizable equations of
the Painlev\'e--Gambier classif\/ication \cite{Pain, Gam, Ince}. The linearizability of the
Gambier equation is based on the fact that it consists in two
Riccati equations in cascade (where by ``cascade'' we mean that
the solution of the f\/irst Riccati equation is used in order to
compute the coef\/f\/icients of the second one). Thus the Gambier
equation, written as a system, has the form:
\begin{subequations}\label{eq1.1}
\begin{gather}
\dot{y}=-y^2+by+c,\label{eq1.1a}\\
 \dot{x} =ax^2+nyx+\sigma.\label{eq1.1b}
 \end{gather}
 \end{subequations}
Here $a$, $b$ and $c$ are functions of the independent variable, $\sigma$ is a constant which can
be scaled to $1$ unless it happens to be $0$ and $n$ is an integer. Eliminating $y$
between \eqref{eq1.1a} and \eqref{eq1.1b} one can write the Gambier equation \cite{GR,GRL} as a second-order equation:
\begin{gather}
\ddot{x}={n-1\over n}{\dot{x}^2\over x}+a{n+2\over n}x\dot{x} +b\dot{x}-{n-2\over n}{\dot{x}\over x}\sigma-{a^2 \over n}x^3 \nonumber\\
\phantom{\ddot{x}=}{} +
(\dot{a} -ab)x^2
+ \Big(cn-{2a\sigma\over n}\Big)x-b\sigma-{\sigma^2\over nx}.\label{eq1.2}
\end{gather}
Equation~\eqref{eq1.2}, although the most general one, is not the only one in the Gambier list. We have in fact
\begin{gather}
\ddot{x}=-3x\dot{x}-x^3+q\big(\dot{x}+x^2\big), \label{eqG5} \tag{G5}\\
 \ddot{x}={\dot{x}^2\over x} +q{\dot{x}\over x}-\dot{q}+rx\dot{x}+\dot{r}x^2,\label{eqG13} \tag{G13}\\
\ddot{x}=\left(1-{1\over n} \right){\dot{x}^2\over x}+qx\dot{x}-
{nq^2\over(n+2)^2}x^3+{n\dot{q}\over n+2}x^2,\label{eqG14} \tag{G14}
\end{gather}
where we have preserved the numbering of Gambier~\cite{Gam}.
To this list of linearizable systems one must in principle adjoin the
equation
\begin{gather}
\ddot{x}=-2x\dot{x}+q\dot{x}+\dot{q}x,\label{eqG6} \tag{G6}
\end{gather}
which is just the derivative of the Riccati equation.

Starting from the Gambier equation one can obtain almost all the
linearizable equations of the Painlev\'e--Gambier list by the appropriate limits.
We start by taking $n=1$ and $\sigma=0$ in~\eqref{eq1.2}. By the appropriate independent variable transformation and gauge we can put $c=0$ and $a=-1$ leading to~\eqref{eqG5}.
Next we start from \eqref{eq1.2} and take the limit $n\to\infty$. We f\/ind
\begin{gather}\label{eq1.3}
\ddot{x} ={ \dot{x}^2\over x}+ax\dot{x} +b\dot{x} -{\dot{x}\over x}\sigma + (\dot{a}-ab) x^2 + dx-b\sigma
\end{gather}
(where we have introduced $d=\lim\limits_{n\rightarrow \infty}cn$).
Equation \eqref{eq1.3} is not in canonical form. We can reduce it to the standard expression by
taking $b=0$  and introducing a gauge $x\rightarrow \rho x$ such that
$d=\ddot{\rho}/\rho- \dot{\rho}^2/\rho^2$. We obtain thus
precisely \eqref{eqG13} where $q=1/\rho$ and $r=a\rho$.
Finally we take $\sigma=0$. and perform an independent variable
transformation~\cite{GRL} which allows to put $b=0$. It turns out that in this case an
additional gauge freedom does exist allowing us to take $c=0$. We
obtain thus the equation
\begin{gather}\label{eq1.4}
\ddot{x} ={n-1\over n}{\dot{x}^2\over x}+a{n+2\over n}x\dot{x} -{a^2 \over n}x^3 + \dot{a}x^2,
\end{gather}
which is precisely \eqref{eqG14} with $q=a(1+2/n)$.

The derivative of the Riccati equation, \eqref{eqG6}, constitutes an exception in the sense that it cannot be obtained from the Gambier equation.
Since we mentioned the derivative of the Riccati it is interesting to point out another ``Riccati derivative'' which is none other but \eqref{eq1.3} or, in canonical form, \eqref{eqG13}. Indeed at the level of the coupled Riccati for the limit $n\to\infty$ to have a meaning $y$ must go to zero as well. This means that the quadratic term in \eqref{eq1.1a} disappears. Moreover we saw that the canonical form corresponds to $b=0$ and since $c$ is free, we can introduce a new function through $d\equiv nc$. Finally, we divide \eqref{eq1.1b} by $x$ and take the derivative. This leads to the appearance of a term $n\dot{y}$, which from \eqref{eq1.1a} is equal to~$d$. Thus, equation \eqref{eq1.4} is a derivative of a Riccati after we have divided by the dependent variable.

At this point one remark is in order concerning the integrability of \eqref{eq1.1}. The Gambier equation was f\/irst derived by Gambier who complemented the study of Painlev\'e of second order dif\/ferential equations the solutions of which have, what is today called, the Painlev\'e property. In order to study the latter in the case of the coupled Riccati system we must start from the observation that since \eqref{eq1.1a} is a Riccati  the dominant behaviour of~$y$  is $y\sim{1/(z-z_0)}$.
An expansion of $y$ around this singularity can be easily established: it involves the functions $b$, $c$ and their derivatives. Next we turn to \eqref{eq1.1b} which is again a Riccati and thus its singularities are simple poles. However
there exist singularities which are due to the singular behaviour of the coef\/f\/icients, namely~$y$.  The
locations of the singularities of the coef\/f\/icients of some equation are considered as `f\/ixed',  i.e.\ not dependent on the initial data. However, in the present case, these singularities are {\it movable} because the $z_0$ in $y$ is arbitrary and thus should be studied. Because of the pole in~$y$, $x$ has a singular expansion. The expansion has a resonance at position $n-1$ and this introduces a compatibility condition. Thus some constraint on the functions $a$, $b$, $c$ must exist for the solutions to have the Painlev\'e property.
However the latter is {\it not} necessary when it comes to integrating a linearisable system. The algebraic integrability which stems from the Painlev\'e property may confer additional ``nice'' properties to the solutions of the system but, if what we are interested in is just to be able to integrate the system, linearizability suf\/f\/ices. In the case of the Gambier equation this means that if one contents oneself with linearizability no constraints on the free functions do exist and as a matter of fact~$n$ does not have to be integer.

\subsection{Motivation and result}

Curiously the Gambier equation has not been the object of many
studies. In particular its relation to other integrable
dif\/ferential systems has not been investigated. This is something
we intend to remedy in the present paper.
We have several results centered around the Gambier equation. These are listed below.
\begin{itemize}\itemsep=0pt
\item We explore a connection between the generalized Ermakov--Pinney \cite{Ermakov, Pinney} and the
Milne--Pinney \cite{Milne,Pinney} equations with the reduced Gambier type equation.

\item Using Okamoto's folding transformation we map the reduced Gambier equation
to an equation of the Li\'enard type. We also derive the conjugate Hamiltonian equations for the
reduced Gambier equation.

\item In addition the conditions under which the reduced Gambier equation
becomes a superintegrable system is investigated.
\end{itemize}

\section{Preliminaries, the generalised Riccati equation}

The properties of higher-order Riccati equations were investigated in \cite{CGR0}
where the authors studied in detail the
Riccati dif\/ferential operator,
\begin{gather*} %\label{E.1}
\D_R=\left(\frac{d}{dt}+h x\right),
\end{gather*}
and the corresponding
Riccati dif\/ferential equations of order $m$ belonging to the
hierarchy
\begin{gather}
\label{E.2} \D_R^m x=0,\quad m=0,1,2,\dots.
\end{gather} Here
 $h$ is assumed to be a constant independent of $t$. Therefore,
as a f\/irst generalization one may assume $h$ to be a
time-dependent function and def\/ine a corresponding dif\/ferential
operator,~$\T_R$, by
\begin{gather*} %\label{E.3}
\T_R=\left(\frac{d}{dt}+h(t)x\right),
\end{gather*}
given in \cite{CGR}.
In a manner similar to
(\ref{E.2}) one may def\/ine, therefore, a nonautonomous hierarchy
of equations by
\begin{gather*}
%\label{E.4}
\T_R^m x=0,\qquad m=0,1,2,\dots.
\end{gather*}
The f\/irst few equations of this hierarchy are given by:
\begin{alignat*}{3}
& n=1\qquad &&\T_R x = \left({\displaystyle{\frac{d}{dt}}} +
h(t) x\right)  x = \dot{x}
+ h(t)  x^2, & \nonumber\\
& n=2 \qquad&&
{\mathbb{T}}_{R}^{2}x=\left({\displaystyle{\frac{d}{dt}}}+
h(t) x\right)^2  x=\ddot{x} +
3 h(t) x \dot{x} + h^2(t)  x^3 + \dot h(t) x^2, & \\ %\label{eq2.5}\\
& n=3\qquad & &{\mathbb{T}}_{R}^{3}x=\left({\displaystyle{\frac{d}{dt}}}+
h(t) x\right)^3  x=\dddot x+4h(t)
x\ddot x + 6h^2(t)  x^2\dot x + 3h(t)  \dot x^2 + h^3(t)  x^4& \nonumber\\
&&& \phantom{{\mathbb{T}}_{R}^{3}x=}{} + 5\dot h(t) x\dot{x} + 3h(t) \dot h(t) x^3 + \ddot h(t) x^2,& \nonumber
\end{alignat*}
and analogous expressions for higher values of $n$. Obviously the
equations become increasingly bulky when $ n > 3$. Of particular
interest is the second member of this family of equations, namely
\begin{gather}
\label{E.6}
\ddot{x}+3h(t)x\dot{x}+\dot{h}(t)x^2+h^2(t)x^3=0.
\end{gather}
 Equation (\ref{E.6}) is called the time-dependent second-order Riccati
 equation and is known to admit a Lagrangian description~ \cite{CGR}. Recently  a time-dependent
 f\/irst integral for this equation has been deduced and this has provided the impetus
 to investigate more deeply the nature of the Riccati operator $\T_R$.

\subsection[Linearization of generalised second-order Riccati equations and the Ermakov-Pinney system]{Linearization of generalised second-order Riccati equations\\ and the Ermakov--Pinney system}

It is well known that the conventional projective transformation
\begin{gather}
\label{ptrans}
 x = \frac{\dot{f}}{hf}
 \end{gather}
   with $h =
$  constant can be applied to linearize the time-independent
second-order Riccati equation
\[
\ddot{x} + 3hx\dot{x} + h^2x^3 = 0.
\]
When $ h = h(t)$ the same projective transformation linearizes the
time-dependent second-order Riccati equation (\ref{E.6}) to the following,
namely,
\begin{gather*}%\label{eq2.8}
 \dddot{f} - \frac{2\dot{h}}{h}\ddot{f} + \bigg( 2\bigg(\frac{\dot{h}}{h}\bigg)^2 - \frac{\ddot{h}}{h}
\bigg)\dot{f} = 0,
\end{gather*}
where $\dot{f}$ denotes dif\/ferentiation with respect to the argument.

Notice that (\ref{E.6}) corresponds to the choice $n=1$, $\sigma = b = 0$ and
$a = - h$ in \eqref{eq1.2}. If $b \neq 0$ then \eqref{eq1.2} reduces (after suitable redef\/inition) of
the time dependent coef\/f\/icients, namely $b \rightarrow -b$, $c\rightarrow -c$
and $a = -h$ to
 \begin{gather*}%\label{eq2.9}
  \ddot{x} + 3hx\dot{x} +
b\dot{x} + cx + \big( \dot{h} + bh \big)x^2 + h^2x^3 = 0.
\end{gather*}
This equation has been considered by Sugai \cite{Su}. Under the
projective  transformation def\/ined by~(\ref{ptrans}) its
corresponding linear version is seen to be
\begin{gather*}%\label{eq2.10}
 \dddot{f} - \bigg( \frac{2\dot{h}}{h} + b \bigg) \ddot{f} + \bigg(2
\bigg(  \frac{\dot{h}}{h}\bigg)^2 - \frac{\ddot{h}}{h} + c -
b\frac{\dot{h}}{h}  \bigg) \dot{f} = 0.
\end{gather*}

\subsubsection[Hill's equation and the Ermakov-Pinney equation]{Hill's equation and the Ermakov--Pinney equation}

We wish to recap \cite{Gu05} here the connection between Hill's equation and the Ermakov--Pinney equation,
which will be generalized in the next section.
It is well known that if $\psi_1$ and $\psi_2$ are the solutions
of the Hill's equation
  \begin{gather*}%\label{eq2.11}
 \Delta \psi  =  \left( \frac{d^2}{dt^2} + u(t) \right)\psi  =  0,
\end{gather*}
then the product $\psi_i \psi_j$   ($i,j = 1,2$) satisf\/ies the third-order linear
equation
\begin{gather}
\label{x1}
\dddot{f} + 2 \dot{u}f + 4u \dot{f} = 0,
\end{gather}
and traces out a three dimensional space of
solutions. Since the solution of equation (\ref{x1}) is spanned by
${\rm Span} ( \psi_{1}^2, \psi_{2}^2, \psi_1 \psi_2)$,  its general
solution  is given by an arbitrary linear combination of the basis
vectors,  viz
\begin{gather*}%\label{eq2.13}
\psi = A\psi_{1}^2 + 2B\psi_1 \psi_2 +
C\psi_{2}^2.
\end{gather*}
One can show that this is in turn connected with
\begin{gather*}%\label{eq2.14}
\ddot{\psi} + u(t) \psi = \frac{\sigma}{\psi^3},
\end{gather*}
where $\sigma$ is a constant.

\begin{proposition}\label{proposition2.1}
If $\psi_1$ and $\psi_2$ satisfy Hill's equation then
\begin{gather*}%\label{eq2.15}
 \psi =
\sqrt{  A\psi_{1}^2 + 2B\psi_1 \psi_2 + C\psi_{2}^2 }
\end{gather*}
satisfies the Ermakov equation
\[
\ddot{\psi} + u(t) \psi = \frac{\sigma}{\psi^3}, \qquad
\sigma = AC - B^2.
\]
\end{proposition}

\begin{proof}
 Since
\[
\psi^2=A\psi_1^2+2b\psi_1\psi_2+C\psi_2^2,
\] dif\/ferentiation
yields
\begin{gather*}
\ddot{\psi} + u(t)\psi =\frac{1}{\psi^3} \Big[\big(A \dot\psi_1^{2} + 2B\dot\psi_1\dot\psi_2 +
C \dot\psi_2^{2}\big)\big(A\psi_1^2+2b\psi_1\psi_2+C\psi_2^2\big) \\
\hphantom{\ddot{\psi} + u(t)\psi =}{} -
\big( A\psi_1 \dot\psi_1 + B(\psi_1 \dot\psi_2+ \dot\psi_1\psi_2 )
+ C\psi_2 \dot\psi_2  \big)^2\Big ].
\end{gather*}
Upon simplif\/ication this gives
\[
\ddot{\psi} + u(t)\psi = \frac{AC-B^2}{\psi^3}W^2.
\] The
proof follows by setting $\sigma=AC-B^2$ and  def\/ining the
Wronskian $W=(\psi_1\dot{\psi_2} - \dot{\psi_1}\psi_2)$ to be
unity.
\end{proof}

It is possible to further extend the above result.

\subsection[Generalized Ermakov-Pinney system and the Gambier equation]{Generalized Ermakov--Pinney system and the Gambier equation}

In this section we extend the previous results.
Conf\/ining ourselves to a special choice of Hill's equation, viz $
u(x)= \Omega^2$, a constant, i.e., the harmonic oscillator equation
we have the following proposition.

\begin{proposition}\label{proposition2.2}
Let  $\psi_1$ and $\psi_2$ satisfy the equation of a linear
harmonic oscillator  then
\begin{gather*}%\label{eq2.16}
 q = \big(A\psi_{1}^{3} +
3B\psi_{1}^{2}\psi_{2} + 3C\psi_{1}\psi_{2}^{2} +
D\psi_{2}^3\big)^{1/3}
\end{gather*}
 satisfies
 \begin{gather}\label{x2}
 \ddot{q} + \Omega^2 q = -\frac{K}{q^5},
 \end{gather}
  where $K$ is some constant and $A$, $B$, $C$, $D$
satisfy following conditions
\[
AC - B^2 = BD - C^2 = -\Lambda \qquad \hbox{and} \qquad AD - BC = 0.
\]
\end{proposition}

\begin{proof}
The proof proceeds in a similar manner as in
Proposition~\ref{proposition2.1}. Since $q^3 = A\psi_{1}^{3} + 3B\psi_{1}^{2}\psi_{2} +
3C\psi_{1}\psi_{2}^{2} + D\psi_{2}^{3}$, where $\ddot{\psi_{i}} =
- \Omega^2\psi_i$ for $i=1,2$,  upon taking the second derivative
we obtain
\begin{gather*}
q^2\ddot{q} + 2q\dot{q}^2 = -\Omega^2\big[A\psi_{1}^{3} + 3B\psi_{1}^{2}\psi_{2} + 3C\psi_{1}\psi_{2}^{2}
+ D\psi_{2}^{3}\big]
\\
\phantom{q^2\ddot{q} + 2q\dot{q}^2 =}{}
+ \big[2A\psi_1\dot{\psi}_{1}^{2} + 2B\dot{\psi}_{1}^{2}\psi_2 + 2C\dot{\psi}_{2}^{2}\psi_1
+ 2D\dot{\psi}_{2}^{2}\psi_2 + 4B\psi_1 \dot{\psi}_{1}\dot{\psi}_{2} + 4C\psi_2 \dot{\psi}_{1}\dot{\psi}_{2} \big].
\end{gather*}
This immediately yields
\begin{gather*}
q^2\big[\ddot{q} + \Omega^2 q\big] = \frac{2}{q^3}\Big[\big( A\psi_{1}^{3} + 3B\psi_{1}^{2}\psi_{2} + 3C\psi_{1}\psi_{2}^{2}
+ D\psi_{2}^{3}\big)
\\
\phantom{q^2\big[\ddot{q} + \Omega^2 q\big] =}{}
\times\big( (A\dot{\psi}_{1}^{2} + 2B\dot{\psi}_{1} \dot{\psi}_{2} + C\dot{\psi}_{2}^{2})\psi_1
+ (B\dot{\psi}_{1}^{2} + 2C\dot{\psi}_{1} \dot{\psi}_{2} + D\dot{\psi}_{2}^{2})\psi_2 \big)\\
\phantom{q^2\big[\ddot{q} + \Omega^2 q\big] =}{}
- \big( (A\psi_{1}^{2} + 2B\psi_1\psi_2 + C\psi_{2}^{2})\dot{\psi}_{2} + (B\psi_{1}^{2} +
2C\psi_1\psi_2 + D\psi_{2}^{2})\dot{\psi}_{2} \big)^2 \Big].
\end{gather*}
Then after a lengthy simplif\/ication process we obtain
\begin{gather*}
\ddot{q} + \Omega^2q
= \frac{2}{q^5}\big[\big(AC-B^2\big)\big( {\dot{\psi}}_1\psi_{2} - \psi_{1}{\dot{\psi}}_2\big)^2\psi_{1}^{2}
+ \big(BD-C^2\big)\big({\dot{\psi}}_1\psi_{2} - \psi_{1}{\dot{\psi}}_2\big)^2\psi_{2}^{2}\\
\phantom{\ddot{q} + \Omega^2q=}{}
 +
(AD-BC)\big({\dot{\psi}}_1\psi_{2} - \psi_{1}{\dot{\psi}}_2\big)^2\psi_1 \psi_2 \big]
\\
\phantom{\ddot{q} + \Omega^2q}{}
= \frac{2}{q^5}\big[ \big(AC - B^2\big)\psi_{1}^{2} + \big(BD - C^2\big)\psi_{2}^{2} + (AD - BC)\psi_{1}\psi_{2} \big]W^2,
\end{gather*}
where $W = ({\dot{\psi}}_1\psi_{2} - \psi_{1}{\dot{\psi}}_2)  $ is
the Wronskian.
Let $AC - B^2 = BD - C^2 = -\Lambda$ and $AD - BC = 0$, then
we obtain
\[
\ddot{q} + \Omega^2q = -\frac{2\Lambda W^2}{q^5} \big(\psi_{1}^{2} + \psi_{2}^{2} \big).
\]
It is known that for harmonic oscillator $\psi_{1}^{2} + \psi_{2}^{2} = E = {\rm const}$.
Thus we obtain
\begin{gather*}
\ddot{q} + \Omega^2q = - \frac{2\Lambda E W^2}{q^5}.\tag*{\qed}
\end{gather*}
\renewcommand{\qed}{}
\end{proof}

The conditions on the coef\/f\/icients as stated above imply that $A + C = B+ D = 0$.

\subsection[Mapping the Ermakov-Pinney equation to the Gambier equation]{Mapping the Ermakov--Pinney equation to the Gambier equation}

Next we map the generalized Ermakov--Pinney equation to a reduced
version of  Gambier's equation. Let
\begin{gather*}%\label{eq2.18}
 x = \frac{a}{q^3},
 \end{gather*}
where $a$ is some constant parameter. By a direct computation one
can show that if $q$ satisf\/ies equation (\ref{x2}) then $x$
satisf\/ies \begin{gather}\label{eq2.19}
 \ddot{x} - \frac{4}{3}\frac{{\dot{x}}^2}{x} -
3\frac{K}{a^2}x^3 -3\Omega^2x = 0.
\end{gather}

Equation \eqref{eq2.19} has a resemblance with the reduced Gambier equation.
This can be easily generalized  for time-dependent $a(t)$.

\begin{proposition}\label{proposition2.3}
Let $x = \frac{a(t)}{q^3}$. If $q$ satisfies the generalized
Ermakov--Pinney equation \eqref{x2} then $x$ satisfies
\begin{gather*}%\label{eq2.20}
 \ddot{x}
- \frac{4}{3}\frac{{\dot{x}}^2}{x}
-\frac{2}{3}\frac{\dot{a}}{a}\dot{x} - 3\frac{K}{a^2}x^3 +
\bigg(\frac{2}{3}\bigg(\frac{\dot{a}}{a}\bigg)^2 -\frac{\ddot{a}}{a} -
3\Omega^2\bigg)x = 0.
\end{gather*}
\end{proposition}

\begin{proof}
 A direct computation of (\ref{x2}) with
$q=(a(t)/x)^{1/3}$ leads to the proof.
\end{proof}

\begin{proposition}\label{proposition2.4}
The equation
\[
\ddot{q}+\lambda \dot{q}+ \mu q = -\frac{A}{q^3}-\frac{B}{q}
\]
 may be mapped to the reduced Gambier equation
\eqref{eq1.2} with $n=-2$ and $\sigma=0$ by the following transformation $x=\alpha/q^2$, where $\alpha$ is a constant.
\end{proposition}

\begin{proof}
The proof follows a direct computation with $q= \sqrt{\alpha}x^{-1/2}$ which gives
\begin{gather*}
\dot{q}=-\frac{\sqrt{\alpha}}{2}x^{-3/2}\dot{x}, \qquad \ddot{q}=-\frac{\sqrt{\alpha}}{2x^{3/2}}\left[\ddot{x}
-\frac{3}{2}\frac{\dot{x}^2}{x}\right].
\end{gather*}
Substituting these expressions into the above equation we obtain
\begin{gather*}
\ddot{x}=\frac{3}{2}\frac{\dot{x}^2}{x}-\lambda \dot{x}+2\mu x+ \frac{2A}{\alpha^2}x^3+\frac{2B}{\alpha}x^2.
\end{gather*}
A comparison allows us to identify the coef\/f\/icients in terms of those occurring in \eqref{eq1.2} with $n=-2$ and $\sigma=0$.
\end{proof}

Note that the transformation $x = \alpha / q^2$ is generally referred to as a folding transformation
and is discussed in the next section.

\subsection{Folding transformations and the Gambier equation}

This section is devoted to the so called {\it folding
transformation} of the Gambier equation. Folding transformations
were f\/irst introduced in \cite{TOS} by Okamoto, Sakai and Tsuda
for  the Painlev\'e equations. According to their terminology
folding transformations are algebraic transformations of the
Painlev\'e systems which give rise to a non-trivial quotient map
of the space of initial conditions. In a more elementary sense one
may say that folding transformations relate the solution of a
given Painlev\'e equation to the square of that of some other
equation (which may be the same as the initial one)~\cite{RGT}.
 One of the interesting results of Okamoto et al.\ was the discovery that P$_{\rm IV}$ possesses a
folding transformation of degree three, mapping solutions into solutions of the same equation
for dif\/ferent values of the parameters.

Let us demonstrate the application of folding transformation for
the Gambier equation. When $b=0$, $n=2$ and $a= {\rm const}$, the
Gambier equation \eqref{eq1.2}
% $$\ddot{x}={n-1\over n}{\dot{x}^2\over x}+a{n+2\over n}x\dot{x} +b\dot{x}-{n-2\over n}{\dot{x}\over x}\sigma-{a^2 \over n}x^3 +
%(\dot{a} -ab)x^2 + \Big(cn-{2a\sigma\over n}\Big)x-b\sigma-{\sigma^2\over nx}, $$
reduces to the following
equation
\begin{gather}
\label{GamR2}\ddot{x}=\frac{1}{2}\frac{\dot{x}^2}{x}+2ax\dot{x}-\frac{a^2}{2}x^3-a\sigma
x-\frac{\sigma^2}{2x}.
\end{gather}
 Under the transformation $x=-q^2$, this
 is mapped to the following equation,
\begin{gather*}%\label{A1}
\ddot{q}+2aq^2\dot{q}+\left\{\left(\frac{a\sigma}{2}\right)q+\left(\frac{a}{2}\right)^2
q^5+\left(\frac{\sigma}{2}\right)^2\frac{1}{q^3}\right\}=0,
\end{gather*}
which is seen to be an equation of the Li\'{e}nard type. On the
other hand if we further set $\sigma=0$ then (\ref{GamR2}) reduces
to
\begin{gather*}%\label{A2}
\ddot{x}=\frac{1}{2}\frac{\dot{x}^2}{x}+2ax\dot{x}-\frac{a^2}{2}x^3,
\end{gather*}
and under the same transformation as above we f\/ind that
\begin{gather*}%\label{A3}
\ddot{q}+2aq^2\dot{q}+ \left(\frac{a}{2}\right)^2
q^5=0.
\end{gather*}
 If however, we set $a=0$ but $\sigma\ne 0$ then
(\ref{GamR2}) assumes the following form
\begin{gather*}
%\label{A4}
\ddot{x}=\frac{\dot{x}^2}{2x}-\frac{\sigma^2}{2x},
\end{gather*}
which under the transformation $x=-q^2$ becomes
\begin{gather*}%\label{eq2.26}
\ddot{q}+\frac{\sigma^2}{4}\frac{1}{q^3}=0.
\end{gather*}

\subsection{Generalized Milne--Pinney system and Gambier equation}

The second-order nonlinear dif\/ferential equation
\begin{gather}
\label{x3}
\ddot{q} + w(t)^2\dot{q} + Kq^{-5} = 0
\end{gather}
 describes the time
evolution of an oscillator with inverse quartic potential. This is
a gene\-ra\-li\-za\-tion of the Milne--Pinney equation where the evolution
is governed by the inverse quadratic potential. It is possible
again to map the generalized Milne--Pinney equation to the Gambier
equation.

\begin{proposition}\label{proposition2.5}
Let $x = \frac{a(t)}{q^3}$. If $q$ satisfies generalized
Milne--Pinney equation \eqref{x3} then $x$ satisfies
\begin{gather*}%\label{eq2.28}
 \ddot{x} -
\frac{4}{3}\frac{{\dot{x}}^2}{x} -
\left(\frac{2}{3}\frac{\dot{a}}{a} -
\frac{w(t)^2}{3a^2}\right)\dot{x} - 3\frac{K}{a^2}x^3 +
\bigg(\frac{2}{3}\left(\frac{\dot{a}}{a}\right)^2 -\frac{\ddot{a}}{a} -
\frac{\dot{a}}{3a^3}w(t)^2\bigg)x = 0.
\end{gather*}
\end{proposition}

\begin{corollary}
Let $w(t)^2 = 2a\dot{a}$. Then the generalized Milne--Pinney
equation is mapped to the reduced Gambier equation
\begin{gather*} %\label{eq2.29}
\ddot{x} -
\frac{4}{3}\frac{{\dot{x}}^2}{x} -\frac{\ddot{a}}{a}x -
3\frac{K}{a^2}x^3= 0.
\end{gather*}
\end{corollary}

\section{Generalized Riccati dif\/ferential operators\\ and Gambier's  equation}

 In this section we construct a generalized Riccati dif\/ferential
 operator which generates a new sequence of dif\/ferential
 equations, that includes the Gambier equation.

We begin by def\/ining the operator
\begin{gather} \label{G.1}
\G
:=\left\{\frac{d}{dt}-\frac{n-1}{n}\frac{\dot{x}}{x}-\left(\frac{\sigma}{nx}
+\frac{b}{2}+\frac{ax}{n}\right)\right\}.
\end{gather}
Here the  dot
represents dif\/ferentiation with respect to $t$ and $n$ is an
integer. Furthermore $a$ and $b$ are assumed to be functions of
$t$ while $\sigma$ is a constant.

  Next we consider the sequence of dif\/ferential equations,
\[
\G^m x=0,\qquad m=0,1,2,\dots,
\]  which give rise to the
 following ODEs:
 \begin{gather*}%\label{G.2}
 m =1, \qquad \G
 x=0\qquad \Rightarrow \qquad \frac{1}{n}\dot{x}=\frac{\sigma}{n}+\frac{b}{2}
 x+\frac{a}{n}x^2,
 \end{gather*}
  which is a generalized Riccati equation.
For $m=2$ we have $\G^2x=0$, which implies
\begin{gather}
\ddot{x}=\frac{n-1}{n}\frac{\dot{x}^2}{x}+a\frac{n+2}{n}x\dot{x}+b\dot{x}-
\frac{n-2}{n}\frac{\dot{x}}{x}\sigma
 -\frac{a^2}{n}x^3\nonumber\\
 \phantom{\ddot{x}=}{} +(\dot{a}-a b)x^2+\left(c
 n-\frac{2a}{n}\sigma\right)x-b\sigma-\frac{\sigma^2}{n x},\label{G.3}
\end{gather}
where
\[
c=\frac{\dot{b}}{2}-\frac{b^2}{4}.
\]
 This procedure can be
carried on leading to higher-order equations.

In this paper we will mostly be concerned with~(\ref{G.3}). As has already been mentioned
(see also~\cite{GRL}) the linearized family of equations belonging to
Gambier's classif\/ication can be obtained as canonical reductions
of this particular equation.

Indeed  for \eqref{eq1.3},  viz
\begin{gather*} %\label{G.4}
\ddot{x}=\frac{n-1}{n}\frac{\dot{x}^2}{x}+a\frac{n+2}{n}x\dot{x}
-\frac{a^2}{n}x^3+\dot{a}x^2,
\end{gather*}
 the corresponding generalized
Riccati operator  is
\begin{gather}
\label{G.5}
\G_{14}=\frac{d}{dt}-\frac{n-1}{n}\frac{\dot{x}}{x}-\frac{a}{n}x.
\end{gather}
This corresponds to the operator for the Gambier equation \eqref{eqG14}.

On the other hand the choice $n=1$, $\sigma=0$, $a=-1$ and $c=0$, so
that  $\dot{b}=b^2/4$, leads to  the f\/ifth equation of Gambier's
classif\/ication scheme, namely
\begin{gather*}%\label{G.6}
\ddot{x}=-3x\dot{x}-x^3+b\big(\dot{x}+x^2\big),
\end{gather*} and corresponds to the
following  generalized Riccati operator
\begin{gather*}%\label{G.7}
\G_{5}=\frac{d}{dt}-\frac{b}{2}-x.
\end{gather*}
  Taking the limit $n\to \infty$ in (\ref{G.3}) we are
  led to the following
  equation,
  \begin{gather*}%\label{G.8}
\ddot{x}=\frac{\dot{x}^2}{x}+ax\dot{x}+b\dot{x}-\frac{\dot{x}}{x}\sigma
  +(\dot{a}-ab)x^2 +dx-b\sigma,
  \end{gather*} where
  $d=\lim\limits_{n\rightarrow\infty}n(\dot{b}/2-b^2/4)$. However, it is
  not quite apparent what the corresponding ge\-ne\-ralized Riccati
  operator for this equation is, since a direct limiting procedure
  on (\ref{G.1}) does not lead to the 13th equation of Gambier's classif\/ication.

\subsection{Hamiltonian framework for the Gambier equation (\ref{eqG14})}

We consider equation~\eqref{eq1.3}
\[
\ddot{x}=\frac{n-1}{n}\frac{\dot{x}^2}{x}+a\frac{n+2}{n}x\dot{x}-\frac{a^2}{n}x^3+\dot{a}x^2.
\]
When $n=-2$ this becomes
\begin{gather}\label{cg1}
\ddot{x}=\frac{3}{2}\frac{\dot{x}^2}{x}+\frac{a^2}{2}x^3+\dot{a}x^2,
\end{gather}
 which may be written in the usual manner as the system
\begin{gather}\label{cgnew}
\dot{x}=y, \qquad \dot{y}=\frac{3}{2}\frac{y^2}{x}+\frac{a^2}{2}x^3+\dot{a}x^2.
\end{gather}
 This system can be expressed in a
 Hamiltonian form with the following Hamiltonian \cite{Cast}:
\[
H(x,p,t)=\frac{1}{2}x^3\left(p+b(t)\right)^2-\dot{a}\log x,
\]
 where $b(t)=\frac{1}{2}\int^t a^2(s) ds$.
 The canonical coordinates used are $q=x$ and
 $p=yx^{-3}-\frac{1}{2}\int^t a^2(s) ds$. The Hamiltonian equations of motion
are
\begin{gather*}
\dot{x} = \frac{\partial H}{\partial p} = x^3 ( p + b ),
\\
\dot{p} = -\frac{\partial H}{\partial x} =  - \frac{3}{2}x^2  ( p + b  )^2 + \frac{\dot{a}}{x}.
\end{gather*}
Using $(p+b) = y/x^3$ we f\/ind that $ \dot{x} = y$. On the other hand
\[
\dot{y} = 3x^2\dot{x} (p+b) + x^3(\dot{p} + \dot{b}).
\]
Since $\dot{b} = 1/2 a^2$ and $\dot{x} = y$, the last expression can be simplif\/ied
to yield the second equation in~(\ref{cgnew}).

\subsection{Conjugate Hamiltonian equation of Gambier type}

Consider a non-autonomous Hamiltonian system in a $(2+1)$-dimensional extended phase space $\{ p, q, t \}$
with a given Hamiltonian $h = H(p, q, t)$. The equation of motion is given by
\begin{gather*}%\label{eq3.10}
\frac{d{ q}}{dt} = \frac{\partial H}{\partial { p}}, \qquad \frac{d{p}}{dt} = -\frac{\partial H}{\partial { q}}.
\end{gather*}
The key point is that the Hamiltonian must be explicitly dependent on time which yields
$t = T({p}, { q}, h)$. Thus
we express
\begin{gather*}%\label{eq3.11}
 h = H(p, q, T( p, q, h))
 \end{gather*}
and by chain rule we obtain
\begin{gather*}%\label{eq3.12}
\frac{\partial T}{\partial { p}} = - \frac{\partial H}{\partial { p}}\left(\frac{\partial H}{\partial t}\right)^{-1}, \qquad
\frac{\partial T}{\partial { q}} = - \frac{\partial H}{\partial { q}}\left(\frac{\partial H}{\partial t}\right)^{-1}, \qquad
\frac{\partial T}{\partial h} = \left(\frac{\partial H}{\partial t}\right)^{-1}.
\end{gather*}
Since $ \frac{dh }{dt} = \frac{\partial H}{\partial t}$
and
\begin{gather*} %\label{eq3.13}
\frac{d{ p}}{dh} = \frac{d{ p}}{dt}\left(\frac{\partial H}{\partial t}\right)^{-1},  \qquad
\frac{d{ q}}{dh} = \frac{d{ q}}{dt}\left(\frac{\partial H}{\partial t}\right)^{-1},\qquad
\frac{dt }{dh}= \left(\frac{\partial H}{\partial t}\right)^{-1}.
\end{gather*}
Substituting \eqref{G.1} in \eqref{G.5} and considering \eqref{G.3} yields
\begin{gather*}%\label{eq3.14}
\frac{d{ p}}{dh} = \frac{\partial T}{\partial { q}}, \qquad \frac{d{ q}}{dh} = -\frac{\partial T}{\partial { p}},
\qquad \frac{dt }{dh} = \frac{\partial T}{\partial h}.
\end{gather*}
This set of equations are known as the {\it conjugate Hamilton's equations}.

\subsubsection{The Li\'{e}nard  equation of second type and its conjugate}

The concept of a conjugate Hamiltonian is introduced in
\cite{Yang}. The solution of the equation $h = H(p,x,t)$, where
$H$ is a given Hamiltonian containing $t$ explicitly, yields the
function $t = T(p,x,h)$. The Hamiltonian system with Hamiltonian
$T$ and independent variable $h$ is said to be  {\it conjugate} to
the initial Hamiltonian system with Hamiltonian $H$. Using this
construction Fokas and Yang~\cite{FY} derived the conjugate
equation for Painlev\'e~II. In particular, they presented a Lax
pair formulation as well as a class of implicit solutions. It may
be mentioned here that, in general, a transformation of the
hodograph type does not preserve the Painlev\'e property. We now
apply their construction to a~Li\'{e}nard equation of second type.

The Hamiltonian of the Lienard  equation of second type,
viz, $ \ddot{x} + f(x) \dot{x}^{2} + g(x) = 0, $ may be
written as
\begin{gather*}%\label{eq3.15}
 H(p,x,t) = \frac{1}{2}e^{-2F(x)}\big(p -
g(x)e^{2F(x)}t \big)^2,
\end{gather*}
 where $F(x) = \int f(x)\, dx$. Thus $H = h$ implies
\[
\pm \sqrt{2}e^{2F(x)} = p - g(x)e^{2F(x)}t.
\]
It can be easily shown that the dual equations are
\begin{gather*}\frac{dx}{dh}  = - \frac{\partial T}{\partial p} =
- \frac{1}{g(x)}e^{-2F(x)},\\ %\label{eq3.16}\\
\frac{dp}{dh}  = \frac{\partial T}{\partial x} =
\frac{d}{dx}\left(\frac{1}{g(x)}e^{-2F(x)}\right)\big(p \mp \sqrt{2h}e^{F(x)}\big) +
\frac{1}{g(x)}e^{-2F(x)}\big( \mp \sqrt{2h}e^{F(x)} f(x) \big). %\label{eq3.17}
\end{gather*}

\paragraph{Conjugate Hamiltonian for the reduced Gambier equation.}
The choice
\begin{gather*}%\label{eq3.18}
f(x) = - \frac{3}{2x} \qquad \hbox{and} \qquad g(x) = -x^3
\end{gather*} leads to the  conjugate of (\ref{cg1}) when
$a(t)=\sqrt{2}$ as given below
 \begin{gather*}%\label{eq3.19}
  \frac{dx}{dh} = 1,
\qquad \frac{dp}{dh} = \mp \frac{3}{2}\sqrt{2h}\,x^{-5/2}.
\end{gather*}
These equations may be solved explicitly yielding
\begin{gather*}%\label{eq3.20}
x = (h + c_1), \qquad p = \mp \frac{3}{\sqrt{2}}\int \sqrt{\frac{h}{(h + c_1)^{5}}}\, dh + c_2,
\end{gather*}
where $c_1$ and $c_2$ are arbitrary constants of integrations.

\section[Generalized Milne-Pinney type equations and superintegrability]{Generalized Milne--Pinney type equations\\ and superintegrability}

 The notion of superintegrability \cite{FMW, WS} has attracted a lot of attention in recent years,
due to its various applications and associated mathematical
structures. Fris et al.~\cite{FMW}, in 1965, studied Euclidean $n
= 2$ systems which admit separability in two dif\/ferent coordinate
systems, and obtained four families of superintegrable potentials
with constants of motion linear or quadratic in the velocities
(momenta). In this section we focus on a certain reduced version
of the  Gambier equation~\eqref{G.3} which exhibits the property of superintegrability. If we set $n =2$ and $a =b = 0$ then equation~\eqref{G.3} reduces to following class of equation
\begin{gather*}%\label{eq4.1}
 \ddot{x} -
\frac{1}{2x}\dot{x}^{2} + \Omega x - \frac{K}{x} = 0,
\end{gather*}  upon
ignoring the constraint on~$c$. Our aim is to study such an
equation using a singular transformation of Sundman type.

  We look for nonlocal transformations, under which a
given ordinary dif\/ferential equation is linearizable. The problem
was studied by Duarte et al.~\cite{Duarte1} by
considering  basically a  transformation of the form
\begin{gather*}%\label{d1}
X(T)=F(t,x),\qquad dT=G(t,x)dt.
\end{gather*} Here $F$ and $G$ are arbitrary
smooth functions  and it is assumed that the Jacobian
$J\equiv\frac{\partial{(T,X)}}{\partial{(t,x)}}\neq 0 $. If one
knows the functional form of $x(t)$, then the latter
transformation ceases to be nonlocal. One of the pioneering
contributors to this f\/ield was K.F.~Sundman~\cite{Sund} who
introduced the transformation $dt = r d\tau$ in the study of the
$3$-body problem, where $r$ is the  dependent variable (radial
component). In particular such transformations are especially
ef\/fective for the solution of several nonlinear ODEs.

  Consider a second-order ordinary dif\/ferential equation
of the Li\'enard-II type
\begin{gather} \label{s.1}
 \ddot{x}+f(x)\dot{x}^2+ g(x)=0,
 \end{gather}
  and suppose we  seek  a nonlocal transformation such
that it is mapped to the  usual  equation of a linear harmonic
oscillator, namely,
 \begin{gather}
  \label{s.2}\ddot{X} + \omega^2 X = 0,
  \end{gather}
   (here
$\dot{X}=\frac{dX}{dt}$) by the nonlocal transformation
\begin{gather*} %\label{s.3}
\frac{dX}{X}=A(x,t)dx+B(x,t)dt, \qquad T = t.
\end{gather*}
 It is a matter of straightforward
computation to show that the SODE (\ref{s.1}) is mapped to
(\ref{s.2}) provided its coef\/f\/icients satisfy the following
conditions:
 \begin{gather}
%\label{s.4}
A^2+A_x - A(x)f(x) = 0,\nonumber\\
%\label{s.5}
B_x+2A(x)B(x) = 0,\nonumber\\
\label{s.6} A(x)g(x) - B^2 = \omega^2.
\end{gather}

Let $A = v_x/v$, then upon
solving the f\/irst and second equations of the above set we obtain
\[
v_x = \exp\left(\int^x f(s)ds\right), \qquad B(x) = \frac{1}{v^2}.
\]
The f\/inal equation \eqref{s.6} then yields
\[
\frac{v_x}{v}g(x) - \frac{1}{v^4} = \omega^2.
\]
In particular, let us set $ f(x) = \frac{\alpha}{x}$, which
readily yields $v = \frac{x^{\alpha + 1}}{\alpha + 1}$ ( $ \alpha
\neq -1$). Then $A$ and $B$ are given by the expressions
\begin{gather*} %\label{eq4.9}
A(x)
= \frac{\alpha + 1}{x} \qquad \mbox{and} \qquad    B(x) =
\frac{(\alpha + 1)^2}{x^{2(\alpha + 1)}}
\end{gather*}
 respectively while
\begin{gather*}%\label{eq4.10}
 g(x) = \frac{(\alpha + 1)^3}{x^{4\alpha + 3}} + \omega^2
\frac{x}{(\alpha + 1)}.
\end{gather*}
 Therefore, the singular Sundman type
transformation is given by
\begin{gather*} %\label{eq4.11}
X = x^{\alpha + 1}\exp\left({(\alpha
+ 1)^2}\int \frac{dt}{x^{2(\alpha + 1)}}\right), \qquad T = t,
\end{gather*}
while  the equation assumes the form
\begin{gather} \label{s.n}
\ddot{x} +
\frac{\alpha}{x}\dot{x}^{2} + \frac{1}{\alpha + 1}\omega^2 x +
\frac{(\alpha + 1)^3}{x^{4\alpha + 3}} = 0.
\end{gather}
 Integrating
(\ref{s.2}) we get
\begin{gather*}%\label{eq4.13}
\left(\frac{dX}{dT}\right)^2 + \omega^2 X^2 =
I(t,x,\dot{x})={\rm const},
\end{gather*} where $I(t,x,\dot{x})$ denotes the
f\/irst integral. Using this recipe we obtain the f\/irst integral of
(\ref{s.n}) as
\begin{gather*}
I = (\alpha + 1)^2 \left( x^{\alpha}\dot{x} + \frac{\alpha + 1}{x^{\alpha + 1}} \right)^2
+ \omega^2 \left( x^{\alpha + 1}\exp\left({(\alpha + 1)^2}\int \frac{dt}{x^{2(\alpha + 1)}}\right) \right)^2.
\end{gather*}
As shown in \cite{AGC1} (and references therein)  the second-order
equation \eqref{s.1} admits a Lagrangian description  via  the
existence of a Jacobi last multiplier (JLM) $ M(x)$, \cite{Jac1} of the form
\begin{gather*}%\label{7.3}
M(x)=e^{2F(x)},\qquad \mbox{where}\qquad F(x)=\int^x
f(s)ds,
\end{gather*}
 with
\begin{gather*}%\label{7.4}
L(x,\dot{x})=\frac{1}{2}M(x)\dot{x}^2
-V(x),\qquad \mbox{where the potential is}\ \ V(x)=\int^x M(s) g(s)
ds.
\end{gather*}
 The following proposition is important in identifying
superintegrable systems.
\begin{proposition}
The necessary condition for mapping of $V(x)$ to
 the superintegrable potential $V(x)=AQ^2+ B/Q^2$ is given by  the condition
 \begin{gather} \label{s.m}
 f(x)=-\frac{5}{4}\left[\frac{g^{\prime\prime}(x)+(f(x)g(x))^\prime}{g^\prime(x)+f(x)g(x)-
2A}\right]
 +\frac{g^{\prime\prime\prime}(x)+(f(x)g(x))^{\prime\prime}}{g^{\prime\prime}(x)+(f(x)g(x))^\prime},
 \end{gather}
  where $Q=\int^x\sqrt{M(s)}ds$ and $A$ and $B$ are arbitrary constants.
\end{proposition}

Clearly one can directly check that (\ref{s.m}) is satisf\/ied by (\ref{s.n}). Moreover
the potential is mapped to
\begin{gather*}
V(x) = \int x^{2\alpha} g(x)\, dx = \int x^{2\alpha}\left[\frac{\omega^2 x}{\alpha + 1} +
\frac{(\alpha + 1)^3}{x^{4\alpha + 3}} \right] dx
 = \frac{1}{2}\frac{\omega^2 (x^{2(\alpha + 1)})}{(\alpha + 1)^2} -
\frac{1}{2}\frac{(\alpha + 1)^2}{x^{2(\alpha + 1)}}.
\end{gather*}
Hence we have $Q=x^{\alpha+1}/(\alpha+1)$ while $A=\omega^2/2$ and
$B=-1/2$.

\section{Conclusion}

In this paper we have examined the Gambier equation and its
associated systems linking them to several well-known dif\/ferential
equations. In particular we have established the simi\-la\-ri\-ty
between special cases of the Gambier equation and the generalized
Ermakov--Pinney and Milne--Pinney systems. The use of Okamoto's
folding transformation leads to a reduction of the Gambier
equation to an equation of the Li\'{e}nard type. In the case of
the Gambier equation~\eqref{eqG14} it is further observed that the system admits a
Hamiltonian description. As a consequence we have  studied the
conjugate Hamiltonian equations for the second type of Li\'{e}nard
equation which resembles the Gambier equation in certain respects
and have indicated their relevance in connection with
superintegrability.

Given that the extension of the Gambier system to the discrete case
has been already established we could speculate on whether some of the
results presented here could be extended to the discrete domain. This
is a question we hope to address in some future work of ours.

\subsection*{Acknowledgements}

We wish to thank  Alfred Ramani for enlightening discussions
and constant encouragements. Two of us (PG and BG) would like to thank
K.M.~Tamizhmani, organizer of the conference {\it Geometry of Integrable Systems}
at Pondicherry, India, for the invitation and his kind hospitality.
One of the authors (AGC) wishes to
acknowledge the support provided by the S.N.~Bose National Centre
for Basic Sciences, Kolkata in the form of an Associateship.

\pdfbookmark[1]{References}{ref}
\LastPageEnding

\end{document}